\documentclass[twocolumn,aps,superscriptaddress,pre]{revtex4}

\usepackage{amsmath,amssymb,graphicx}
\usepackage{algorithmic}
\usepackage{enumerate}
\usepackage{color}
\usepackage{soul,xcolor}  
\setstcolor{red} 

\graphicspath{{./fig-jpg/}{./fig-ps/}}

\begin{document}

\title{Dynamics of interacting dark soliton stripes}

\author{P. G. Kevrekidis}
\email{kevrekid@math.umass.edu}
\affiliation{Department of Mathematics and Statistics, University of Massachusetts,
Amherst, Massachusetts 01003-4515 USA}

\author{Wenlong Wang}
\email{wenlongcmp@gmail.com}
\affiliation{Department of Theoretical Physics, Royal Institute of Technology, Stockholm, SE-106 91, Sweden}

\author{G. Theocharis}
\affiliation{Laboratoire d'Acoustique de l'Universit\'{e} du Maine, UMR-CNRS 6613, Av.~Olivier Messiaen, Le Mans 72000, France}

\author{D. J. Frantzeskakis}
\affiliation{Department of Physics, National and Kapodistrian University of Athens,
Panepistimiopolis, Zografos, 15784 Athens, Greece}

\author{R. Carretero-Gonz{\'a}lez}
\affiliation{Nonlinear Dynamical Systems
Group,\footnote{\texttt{URL}: http://nlds.sdsu.edu}
Computational Sciences Research Center, and
Department of Mathematics and Statistics,
San Diego State University, San Diego, California 92182-7720, USA}

\author{B. P.~Anderson}
\affiliation{College of Optical Sciences,
University of Arizona, Tucson, AZ, 85721, USA.}

\begin{abstract}

In the present work we examine the statics and dynamics of multiple parallel 
dark soliton stripes in a two-dimensional Bose-Einstein condensate. Our principal 
goal is to study the effect of the interaction between the stripes on the 
transverse instability of the individual stripes. We use a recently 
developed adiabatic invariant formulation to derive a quasi-analytical 
prediction for the stripe equilibrium position and for the  
Bogoliubov-de Gennes spectrum of excitations of stationary stripes. 
The cases of two-, three- and four-stripe states are studied in detail.
We subsequently test our predictions against numerical 
simulations of the full two-dimensional Gross-Pitaevskii equation. 
We find that the number of unstable eigenmodes increases as the number 
of stripes increases due to (unstable) relative motions between the stripes.
Their corresponding growth rates do not significantly change, although
for large chemical potentials, the larger the stripe number, the larger
the maximal instability growth rate. 
The instability induced dynamics of multiple stripe states and their 
decay into vortices are also investigated.

\end{abstract}

\maketitle

\section{Introduction}

The study of dark solitons has attracted considerable theoretical and experimental 
attention in various branches of physics. Prominent examples include
nonlinear optics~\cite{Kivshar-LutherDavies} and 
atomic Bose-Einstein condensates (BECs)~\cite{djf,SIAMbook}, 
but also mechanical \cite{mech} and electrical \cite{el} dynamical lattices, 
magnetic films \cite{mf}, electromagnetic \cite{gv} and acoustic \cite{acous} metamaterials, 
hydrodynamics \cite{ww}, plasmas \cite{plasma}, nematic liquid crystals \cite{nlc}, 
as well as other nonlocal media \cite{nonloc}, dipolar atomic condensates \cite{chrom}, 
and exciton-polariton condensates \cite{polar}.

In the context of BECs, a number of experiments have addressed 
diverse phenomena including the formation of dark solitons by  
laser beams dragged through an elongated BEC~\cite{engels}, 
their oscillations in the trap~\cite{becker,ourmarkus2}, their 
pairwise interactions~\cite{ourmarkus2,ourmarkus3}, as well as their 
transverse instability in higher-dimensional settings 
and their concomitant decay into vortices and vortex
rings~\cite{watching,jeff,becker2,lamporesi} (see the recent volume~\cite{SIAMbook} 
summarizing many of these phenomena). Furthermore,
a sizeable literature has emerged in the topic of multi-component 
condensates, where one of the components assumes the form of a dark soliton 
(see, e.g., the recent review~\cite{revip} and references therein). 
Regarding applications, matter-wave dark solitons have been proposed for use in 
atomic matter-wave interferometers~\cite{in1} and as qubits in BECs~\cite{qub}.

Generally, dark solitons are unstable against decay when embedded in a higher dimensional space.
The study of their transverse (or ``snaking'') instability 
has been of particular significance since its theoretical inception~\cite{kuzne} 
(see also the review~\cite{kidep} and references therein). 
However, the instabilities associated with the mutual interactions between multiple dark 
solitons in a BEC 
have only been partly investigated ---see, e.g., Ref.~\cite{ourmarkus3} for the 
quasi-1D setting, as well as the recent work~\cite{mat} for 2D and 3D settings.
Importantly, these works have always considered the role of trap
induced confinement in the corresponding dimensionality of the problem.
In the higher-dimensional cases, examples of stability have been
reported numerically
for suitable parametric regimes suppressing the snaking instability.

In the present work, we analytically and numerically characterize the dynamics of up 
to four parallel dark solitons, finding that while the number of unstable eigenmodes 
increases as the number of stripes increases, their corresponding growth rates do not 
significantly change; nevertheless for large chemical potentials the rates
are found to increase with the stripe number.  Progress on this front is especially relevant in experimental settings that involve solitons produced in the merging or collisions of BECs~\cite{kali}, the interaction of multiple dark solitons~\cite{ourmarkus2,ourmarkus3,jeff,becker2}, or the decay of multiple dark solitons as a generator of vortices~\cite{watching} and two-dimensional (2D) quantum turbulence~\cite{tsubota} in BECs as well as in optics~\cite{tikh}.
The subject of dark soliton decay also continues to attract theoretical attention~\cite{smirnov,hoefer}, including the possibility of ``engineering'' avoidance of this instability~\cite{us}.


In recent work~\cite{aipaper,aipaper2}, we provided a framework 
for addressing the transverse instability of a diverse array of structures 
including dark soliton stripes, ring dark solitons (extensively studied in 
optics~\cite{kivyang,rings,rings1} and BECs~\cite{rings2,rings3,rings4,korneev}), 
and spherical shell solitons~\cite{kivyang,carr,wenlong,hau}, as well as 
dark-bright solitons in multi-component BECs~\cite{revip}.    
Our approach in the present work involves extending this formulation to the case of multi-soliton settings. 
%
We seek to understand how the presence of a secondary
stripe may affect the growth rate of a transverse instability.
To address this problem we combine the adiabatic invariant (AI) of
a single dark soliton stripe~\cite{aipaper2} with the pairwise
interaction between the stripes. This allows us to identify
the equilibrium states of the multiple stripes, and more importantly the
modes of linearization as per the well established
Bogolyubov-de Gennes (BdG) analysis, 
around the solitonic multi-stripe solution. Finally, this formulation
enables an exploration of the fully nonlinear stage of the instability
by examining the filament partial differential equation (PDE) for each
of the relevant stripes. While the analytical framework becomes
rather complex as the number of stripes increases, we explore
the relevant instability numerically for larger stripe numbers
(such as 3 and 4).

Our presentation is structured as follows. In Sec.~\ref{theory} 
we provide the theoretical analysis of the case of two dark soliton stripes, which we refer to as ``2-stripes.'' In Sec.~\ref{results}, we explore numerically the scenarios of 2-, 3- and 4-stripes 
and, where appropriate, compare with the semi-analytical predictions 
of the filament theory. Finally, in Sec.~\ref{cc}, we summarize our 
findings and present our conclusions, as well as a number of directions for future work.

\section{Theoretical Analysis}
\label{theory}

The model under consideration is the normalized 2D Gross-Pitaevskii 
equation (GPE),
describing a condensate confined in a highly oblate trap
along the $z$-axis of frequency $\omega_z$~\cite{SIAMbook}:
%
  \begin{eqnarray}
    i u_t =-\frac{1}{2} \left(u_{xx}+u_{yy}\right) + |u|^2 u + V(x) u,
    \label{extr}
  \end{eqnarray}
where subscripts denote partial derivatives, and $u(x,y,t)$ denotes the wavefunction.
Here, the density $|u|^2$, length, time, and energy are respectively measured in units of 
$2\sqrt{2\pi}a a_z$, $a_z$, $\omega_z^{-1}$, and
$\hbar\omega_z$, where $a$ and $a_z$ are, respectively, the $s$-wave scattering 
length and harmonic oscillator 
length in $z$-direction. The external potential is given by
\begin{equation}
V(x)=\frac{1}{2} \Omega^2 x^2,
\label{htrap2dd}
\end{equation}
which is independent of the transverse $y$-direction, with $\Omega = \omega_x/\omega_z$ 
being the trap's aspect ratio.
After this dimension reduction, we carry out a subsequent scaling and consider $\Omega=1$ (see Sec.~\ref{bg} for more details).

The model is supplemented with periodic boundary conditions in the $y$-direction. 
Equation~(\ref{extr}), for $V=0$, conserves the Hamiltonian:
\begin{eqnarray}
H=\frac{1}{2}  \iint_{-\infty}^{\infty} 
  \left[ |u_x|^2 + |u_y|^2 +
  \left(|u|^2-\mu \right)^2 \right] dx\, dy, 
\label{ham}
\end{eqnarray}
where $\mu$ is the chemical potential. In the dimensionless form of the GPE given here, 
we consider chemical potentials ranging from the linear limit up to $\mu$=80 
in the Thomas-Fermi limit.
%
This range is sufficient to address the chemical potential and atomic densities in typical experimental BECs; see Ref.~\cite{SIAMbook} for a detailed
discussion on the translation between dimensionless and dimensional
units.

Let us first study the case of two-dark soliton stripes. To describe each stripe,
we consider the following ansatz: 
\begin{eqnarray}
u=e^{-i \mu t} \left[
  \sqrt{\mu - v^2} \tanh \left(\sqrt{\mu - v^2} (x-x_0) \right) + i v \right],
\label{dark2a}
\end{eqnarray}
which is the functional form of the quasi-one-dimensional dark soliton solution of Eq.~(\ref{extr}),  
for $V=0$, that extends uniformly in the $y$-direction. In the 2D setting under 
consideration, Eq.~(\ref{dark2a}) describes a one-dimensional
dark soliton stripe embedded in 2D space, 
characterized by its center, $x_0$, and velocity $v={x}_{0t} \equiv dx_0/dt$.
In order to describe the transverse instability-induced undulation of the stripe, 
we assume that the center position $x_0$ is not solely a function of $t$, but also 
a function of the transverse variable $y$, i.e., $x_0=x_0(y,t)$.
We also wish to consider cases in which the potential $V(x)$ may be non-zero, which involves replacing $\mu$ with $\mu-V(x_0)$ ---representing the effective, or {\em local}, chemical
potential where the dark soliton is sitting--- in Eqs.~(\ref{dark2a}) and (\ref{ham}). 
Then, substituting 
the ansatz~(\ref{dark2a}) into the Hamiltonian~(\ref{ham}), we obtain the following
``effective energy'' (an adiabatic invariant) of the stripe:
\begin{eqnarray}
  E= \frac{4}{3} \int_{-\infty}^{\infty}  
  \left(1 + \frac{1}{2}x_{0y}^2 \right)
  \left(\mu - V(x_0) -x_{0t}^2 \right)^{3/2}  dy.
  \label{dark5}
\end{eqnarray}
Here, the transverse energy contribution (of the $|u_y|^2$ term) has
been accounted for through the term proportional to $x_{0y}^2$.
For convenience, hereafter we use the following compact notation:
$$A=\mu - V(x_0) - x_{0t}^2, \quad B= 1 + \frac{1}{2}x_{0y}^2.$$

Earlier work on the interaction of dark solitons~\cite{akh,kk} in one-dimensional (1D) settings, 
including relevant work in the context of quasi-1D atomic BECs~\cite{ourmarkus3,ourmarkus2,coles}, has quantified the interaction 
effect between dark solitons. Now, this interaction
becomes a pointwise effect across the stripes, when the interaction
term is integrated along $x$, as per the variational formulation of Ref.~\cite{kk}.
Thus the corresponding energy, incorporating through its last term this interaction
effect, reads:
  \begin{eqnarray}
    E=2 \int_{-\infty}^{\infty}
    \left( \frac{4}{3} A^{3/2} B - 8 A^{3/2} e^{-4 A^{1/2} x_0} \right) dy.
\nonumber
  \end{eqnarray}
Here, we have used the symmetry of the two solitons, which are assumed to 
be located at $\pm x_0(y)$. This represents the simplest possible scenario, 
where a single dynamical variable, $x_0$ (i.e., the symmetric position of the two solitons),  
can adequately describe the dynamics of both. Nevertheless, it should be pointed out that 
this scenario is of direct relevance to experiments~\cite{ourmarkus3,ourmarkus2}. 
Due to the consideration of two solitons, the energy of each individual
soliton is doubled, hence the factor of $2$ in front of the integral.
  
We can now find the evolution of the 2-stripe parameter $x_0$ from 
the energy conservation, $dE/dt=0$. Indeed, the relevant calculation leads 
to the following PDE for $x_0$:
  \begin{eqnarray}
    B \left( {x_0}_{tt} + \frac{V'}{2}\right) + \frac{A}{3} {x_0}_{yy}
    = \frac{V'}{2} {x_0}_y^2 + {x_0}_y {x_0}_t {x_0}_{ty} 
\nonumber
\\[1.0ex]
\label{extr2}
     -\left[(V'+ 2 {x_0}_{tt}) (-3 + 4 A^{1/2} x_0) - 8 A^{3/2} \right]
    e^{-4 A^{1/2} x_0},
  \end{eqnarray}
where $V'\equiv \partial V/\partial x_0$. Hereafter, the above equation 
will be called the ``adiabatic invariant PDE'' (AI PDE).

The next step is to examine the stationary states of this PDE and their stability. 
As we know, there exists a homogeneous (independent of $y$) solution
corresponding to the two parallel 
soliton stripes. Retrieving that, as well as its linearization, yields information 
about the existence and stability of the 2-stripe state, and more specifically on how
the presence of a second stripe affects the transverse (in)stability of the first one.

More specifically, we seek a symmetric pair of soliton stripes with $x_0$ 
independent of $y$ that yields the following algebraic (transcendental)
equation for $x_0$ (see, e.g., Ref.~\cite{ourmarkus3} for a relevant analysis):
  \begin{eqnarray}
    -\frac{V'}{2}=e^{-4 A_0^{1/2} x_0}
    \left[V'(-3 + 4 A_0^{1/2} x_0) - 8 A_0^{3/2} \right],
    \label{extr4}
  \end{eqnarray}
where $A_0=\mu-V(x_0)$.
Notice that results stemming from Eq.~(\ref{extr4}) are expected to be more accurate 
in the framework of the so-called ``particle approximation'', whereby individual 
solitons feature a particle-like nature; this situation corresponds to the case of 
sufficiently large values of the chemical potential $\mu$, the so-called Thomas-Fermi (TF)
large density limit. 
Thus, this step leads to the determination of the equilibrium positions $\pm x_0^{(\rm{eq})}$ 
of the constituent stripe solitons forming the stationary stripe pair. Then,  
the stability of $x_0^{(\rm{eq})}$ can be studied by introducing the ansatz
\begin{equation}
x_0=x_0^{(\rm{eq})} + \epsilon X_1(t) \cos(k_n y),
\nonumber
\end{equation}
with $k_n=n\pi/L_y$ denoting the transverse perturbation wavenumbers, and $L_y$ being 
the size of the computational domain in the $y$-direction ---see below. Then, we 
linearize with respect to the small-amplitude perturbation $X_1(t)$, and  
determine whether such a stationary stripe pair is robust under transverse 
modulations or not. By doing so, we obtain a rather elaborate expression that 
can be summarized as
  \begin{eqnarray}
    R X_{1tt}= - \left[ \frac{1}{2}V''(x_0) - \frac{1}{3}k_n^2 A_0 + S \right] X_1.
\nonumber
    \end{eqnarray}
Here, $A_0=A(t=0)$ and the coefficients $R$ and $S$ are given by: 
\begin{eqnarray}
R&=&1 +2 (-3 + 4 A_0^{1/2} x_0) e^{-4 A_0^{1/2} x_0}, 
\nonumber
\\[2.0ex]
\nonumber
S&=& R \left[-4 S_1 \left( V'(x_0) S_2-8 A_0^{3/2}\right)+ 4V'(x_0) S_1 \right]
\\
\nonumber
      &&+ \left[V''(x_0) S_2
      +12 V'(x_0) A_0^{1/2}\right] R_0,
\end{eqnarray}
where $S_0=e^{-4 A_0^{1/2} x_0}$, $S_1=A_0^{1/2} - V' A_0^{-1/2} x_0/2$, and 
$S_2=-3 + 4 A_0^{1/2} x_0$.
%
%
%
Note that here, for simplicity, we have used $x_0$ instead $x_0^{(\rm{eq})}$; 
furthermore, both $R$ and $S$ are evaluated at the equilibrium position $x_0=x_0^{(\rm{eq})}$. With 
the equation of motion for a generic perturbation $X_1$, we can look for 
normal mode frequencies by setting 
$$X_1(t) = X_1^{(0)} \exp(\lambda t),$$ 
where $X_1^{(0)}$ is a constant amplitude, 
and $\lambda$ is the eigenvalue of the normal mode $n$.
  
In the numerical section below, we compare this prediction about multi-stripe 
stability with detailed computations of the spectrum at the 2D GPE level. This 
is done upon performing the standard Bogoliubov-de Gennes (BdG) analysis (see details, e.g., 
in Ref.~\cite{SIAMbook}). Moreover, given the favorable comparison reflected in our results,  
it is relevant to examine the full AI PDE dynamics 
of Eq.~(\ref{extr2}) against the corresponding 2D dynamical evolution 
of the GPE. This will help us to build a systematic appreciation of 
the role of soliton interactions in enhancing or mitigating transverse 
instabilities. It will also serve as a stepping stone towards generalizing 
this to a larger number of stripes, by means of the (AI PDE) extension of the 
significantly simpler, ODE-based, 1D picture of Refs.~\cite{ourmarkus3,coles}.

\section{Numerical results}
\label{results}



\subsection{Preliminaries}
\label{bg}

It is worth noting that the theoretical analysis in the last section applies 
to a generic potential $V(x)$ varying slowly on the soliton scale. In
what follows, we focus on the experimentally relevant harmonic trap described by 
Eq.~(\ref{htrap2dd}). 
For all of the numerical computations we set the trap strength $\Omega=1$.
However, it is important to mention that,
after a scaling transformation, smaller values for $\Omega$ would 
correspond to larger values of the chemical potential $\mu$.
Specifically, by defining the rescaled variables 
$(\bar{x},\bar{y})=\bar{\Omega}^{-1/2}\,(x,y)$, $\bar{u}=\bar{\Omega}^{1/2}\,\,u$, 
and $\bar{\mu}={\bar{\Omega}}\,\mu$, 
Eq.~(\ref{extr}) with $\Omega=1$ transforms to the same equation but with
arbitrary $\bar{\Omega}$.

As for our computational domain we use $(x,y)\in[-L_x,L_x]\times[-L_y,L_y]$ with 
periodic boundary conditions in $y$. 
The values of $L_x$ and $L_y$ 
will be chosen as $L_x=16$ and $L_y=2$, $L_y=4$, or $L_y=8$. In terms of
physical
parameters, and for $L_y=4$, one may estimate the system parameters corresponding to a $^{87}$Rb 
condensate confined in a 
parabolic trap with $\omega_z=2\pi\times 100$~Hz. In this case, the number of atoms 
(resulting from the normalization of the wavefunction) is given by: 
$$N=\frac{4L_y}{3\sqrt{\pi}} \left(\frac{a}{a_z} \right)\mu^{3/2}.$$
For $^{87}$Rb, the $s$-wave scattering length is $a=5.3$~nm. The 
trapping frequency $\omega_z$ corresponds to a harmonic oscillator 
length $a_z=\sqrt{\hbar/(m\omega_z)}=2.7~\mu{\rm m}$.
With these parameters, typical numbers of atoms, as well 
as respective values of the healing length $\xi=\hbar/\sqrt{2m g_{2D} n_{2D}}$ 
(where $g_{2D}=2\sqrt{2\pi} a a_z\hbar \omega_z$ is the effective 2D interaction 
strength and $n_{2D}$ the peak density \cite{SIAMbook}), are as follows: 
for a (dimensionless) chemical potential 
$\mu=1$, one has $N\sim1400$ atoms and $\xi \sim 1.9~\mu{\rm m}$;
for $\mu=40$ the respective values are $N \sim 3.5 \times 10^5$ atoms and $\xi \sim 0.3~\mu{\rm m}$, 
while for $\mu=80$, we have $N \sim 10^6$ atoms and $\xi \sim 0.2~\mu{\rm m}$.
Notice that the above characteristic values of the chemical potential 
correspond to various numerical results that will be presented below. Furthermore,
in the simulations depicting stripe dynamics, the time unit $t=1$ corresponds to $\sim 2$~ms. 

With respect to our numerical simulations, we 
use a computational framework, similar to our earlier
works~\cite{aipaper,aipaper2,wenlong}. This is based
on finite differences to approximate partial derivatives and the 
partial wave method to reconstruct the spectrum for the full (2D) 
system by computing a handful of 1D ($x$-) spectra.
To fully resolve the 1D spectra we used 64,000 mesh points
over the interval $-16\leq x \leq 16$.

\subsection{Two dark soliton stripes}

\begin{figure}[tbp]
\begin{center}
\includegraphics[width=\columnwidth]{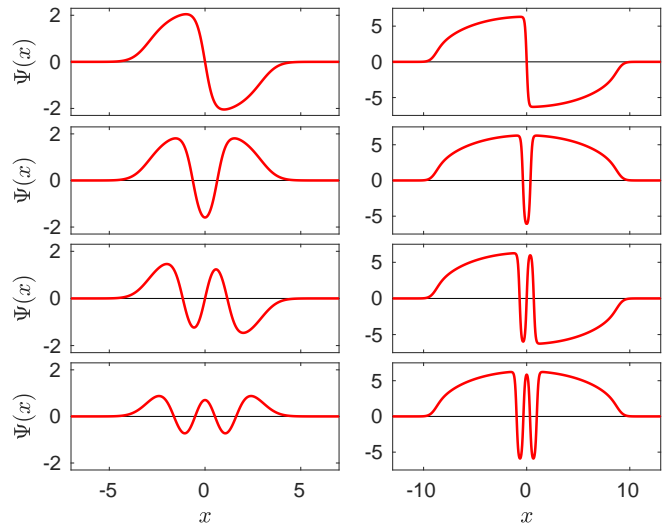}
\caption{(Color online) 
Cross sections ($y={\rm const.}$) along the $x$-direction of typical wavefunctions 
$\Psi$ [which are found from the GPE~(\ref{extr}) using 
$u(x,y,t)=\Psi(x) \exp(-i\mu t)$]
corresponding from top to bottom to one, two, three and four dark solitons, 
respectively. In all cases $\Omega=1$ and $\mu=5$ (left) and $\mu=40$ (right)
correspond to chemical potential values close to the linear (small density) 
and Thomas-Fermi (large density) limits, respectively. 
Note that these 2D stationary states are homogeneous in the $y$-direction, 
as the potential (\ref{htrap2dd}) is only $x$-dependent.
}
\label{States}
\end{center}
\end{figure}

In the linear, low-density limit of Eq.~(\ref{extr}), each of the dark
soliton multi-stripe states corresponds to an eigenstate of the simple harmonic 
oscillator. That is, at chemical potential $\mu=(n+1/2)\Omega$, 
the $n$-th harmonic oscillator eigenfunction is a starting point for
the continuation of a nonlinear state with $n$ stripes (see, e.g., Ref.~\cite{alf}). We thus
use parametric continuation towards higher values of the chemical
potential, gradually tending to the large chemical potential limit.
In Fig.~\ref{States} we depict the different steady states with one
to four stripes for the small and large density limits.
It is important to stress that it is in the Thomas-Fermi (large density) limit that 
we expect the above developed theory to be valid. This is because, in 
that limit, the dark-soliton stripe width $\propto \mu^{-1/2}$
tends to $0$, thus justifying its consideration as a filament (without
internal dynamics).

\begin{figure}[tbp]
\begin{center}
\includegraphics[width=\columnwidth]{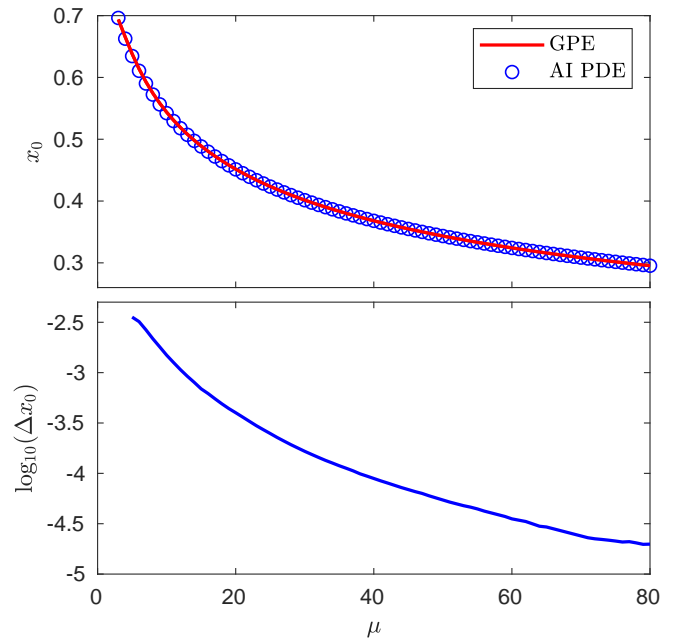}
\caption{(Color online) 
Top: Equilibrium position $x_0=x_0^{({\rm eq})}$ for the 2-stripe, as 
obtained from the GPE~(\ref{extr}) (red line) 
and the AI PDE model~(\ref{extr2}) (blue circles). 
Bottom: Difference $\Delta x_0$ between the equilibrium position from the 
GPE~(\ref{extr}) and the one predicted by AI PDE model~(\ref{extr2}).
Notice the remarkably good agreement even near the linear limit.
}
\label{fig:EquiDS2}
\end{center}
\end{figure}

We start with the equilibrium points of the (homogeneous) stationary 
stripes. Here, the comparison of the analytical prediction of Eq.~(\ref{extr4}) with 
the full numerical results of the GPE is depicted in Fig.~\ref{fig:EquiDS2}. 
A remarkably good agreement is found between the two, for all of the considered values 
of the chemical potential, with the result indeed becoming essentially exact 
for sufficiently large values of $\mu$ (five digits of precision for $\mu \approx 80$).
In some sense, however, this agreement 
should be expected, given the corresponding 1D results of Ref.~\cite{coles}, 
as well as the effectively quasi-1D nature of the pertinent equilibria. 
A far more challenging test of the theory lies in the examination of the 
corresponding linearized modes.

\begin{figure}[tbp]
\begin{center}
\includegraphics[width=0.90\columnwidth]{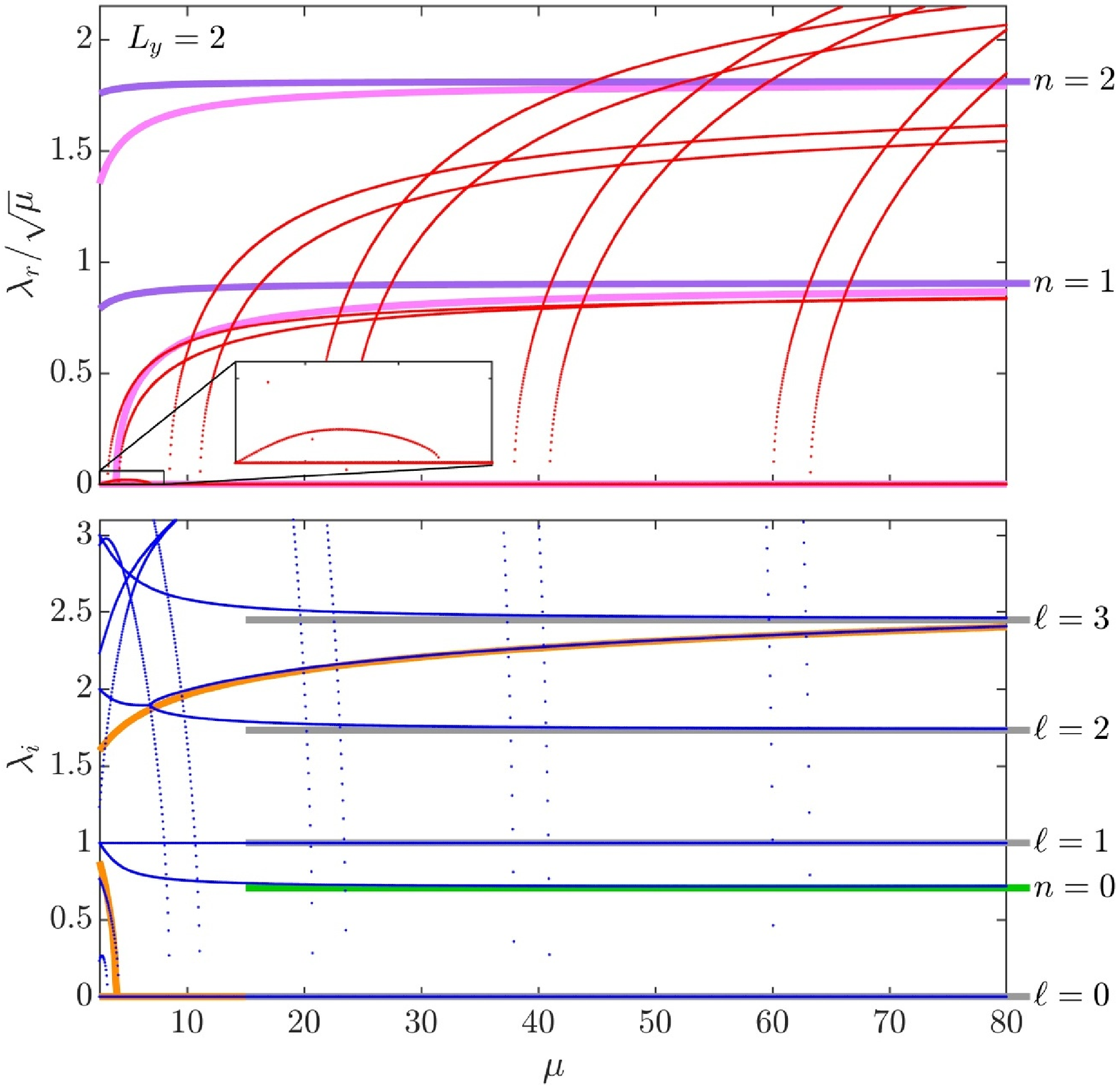}
\\[2.0ex]
\includegraphics[width=0.90\columnwidth]{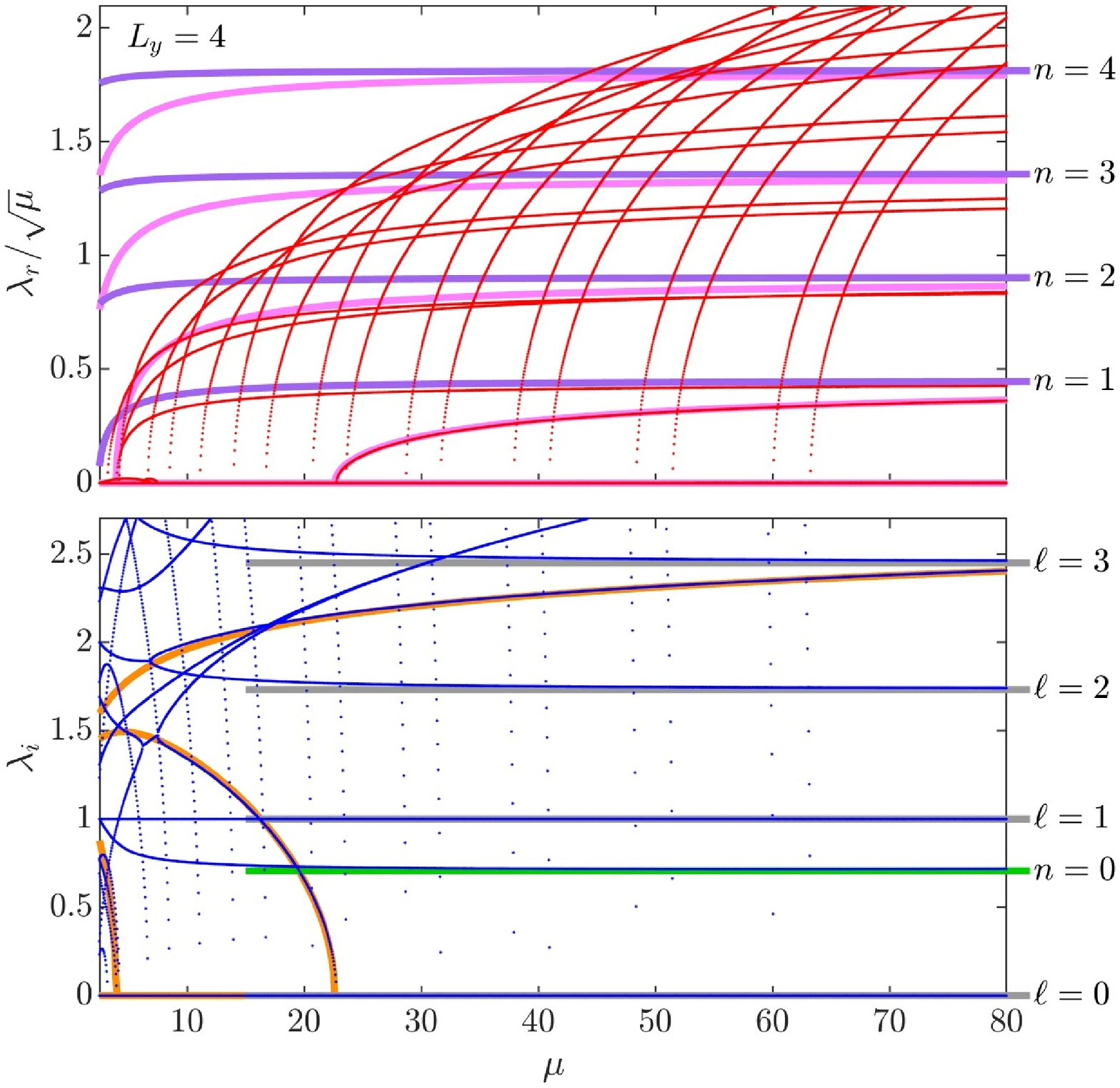}
\caption{(Color online) 
Comparison between the two-dark soliton stripes' stability spectra 
for the full GPE and the analytical prediction based on the AI PDE.
Depicted are the real ($\lambda_r$; top subpanels) and imaginary 
($\lambda_i$; bottom subpanels) parts of the stability eigenvalues, 
$\lambda=\lambda_r + i\,\lambda_i$, vs.~the chemical potential $\mu$ 
(note that $\lambda_r$ is scaled by $\sqrt{\mu}$). 
For the numerical domain we use $L_x=16$, as well as $L_y=2$ (top
set of panels) and $L_y=4$ (bottom set of panels). 
Red and blue dots correspond to the real and imaginary parts of the spectrum
from the full GPE model, while pink and orange thick curves correspond 
to the real and imaginary parts for the effective AI PDE model.
The thick green horizontal line is the single-stripe $n=0$ (stable) in-phase mode 
Im$(\lambda)=\Omega/\sqrt{2}$ while the thick violet curves depict the (unstable)
$n\geq1$ in-phase modes; see Eq.~(\ref{eq:IPkn}).
The thick gray horizontal lines correspond to the 1D TF spectrum; see 
Eq.~(\ref{eq:TFspectrum}).
%
}
\label{DS2spectrum}
\end{center}
\end{figure}

Once we have identified the steady states for $N$ dark soliton stripes,
we proceed to compare their respective spectra of the BdG analysis of  
the GPE (``BdG/GPE'' hereafter) with the corresponding predictions 
of the AI PDE, Eq.~(\ref{extr2}). 
At this point, it should be recalled that, for mathematical simplicity, 
we have restricted consideration to two dark soliton stripes symmetrically 
placed around the center of the trap (in line with, e.g., the experiments 
of Refs.~\cite{ourmarkus2,ourmarkus3}). Therefore, the AI theory will 
naturally capture only the normal modes corresponding to dark soliton stripes 
oscillating {\it out-of-phase} (OOP); for simplicity these modes will 
be referred hereafter to as OOP modes. 
Figure~\ref{DS2spectrum} depicts the spectra obtained via 
BdG/GPE [(red and blue) dots] and AI PDE [thick (pink and orange) curves] for the
2-stripe steady state. It is relevant to mention that
the spectra have a strong dependence on the numerical domain, 
namely $(x,y)\in[-L_x,L_x]\times[-L_y,L_y]$, and particularly 
on $L_y$. 
Specifically, since the snaking instability of dark soliton stripes
is only present for relatively small wavenumbers, reducing the length
of the domain in the $y$-direction results in long wavelengths being
suppressed (i.e., not accessible) on this smaller domain. As a result, 
as $L_y$ is reduced, more (long-wave) modes are suppressed and only
shorter wavelength eigenmodes will be unstable (cf.~compare the $L_y=2$
and the $L_y=4$ spectra in Fig.~\ref{DS2spectrum}). In fact, for
small enough $L_y$, no unstable wavelengths will fit in the domain 
and the solutions will be effectively stable in a manner akin to the
stabilization of transverse dark solitons in elongated BECs
reported in Ref.~\cite{mat}.
Therefore, we have selected two typical $L_y$ values to compare 
the BdG/GPE and AI PDE. The results for $L_y=2$ and $L_y=4$ are shown 
in Fig.~\ref{DS2spectrum} (see top and bottom sets of panels, respectively).
As can be seen from the figure, all of the AI PDE modes referring to the soliton stripes 
are also present in the original BdG/GPE. 

It is crucial to note that the AI PDE 
cannot capture modes referring to the background (stripe-less configuration). 
These background modes correspond to both longitudinal and transverse 
modes. The longitudinal background modes in our system correspond to the collective 
oscillations of a 1D trapped BEC, characterized by the imaginary eigenvalues 
$\lambda_\ell=i\omega_\ell$, where the corresponding frequencies are given by \cite{menstr}:
\begin{equation}
\label{eq:TFspectrum}
\omega_\ell = \sqrt{\frac{\ell(\ell+1)}{2}}\,\Omega.
\end{equation}
Note that in our notation, instabilities are characterized by a positive real
part of the eigenvalue $\lambda$, which in turn corresponds to an imaginary part
on the corresponding eigenfrequency $\omega$.
The frequencies of the longitudinal background modes are depicted by the thick gray 
horizontal lines in Fig.~\ref{DS2spectrum}, which give a good approximation
of the corresponding frequencies of the full BdG/GPE modes 
---especially as $\mu$ becomes larger.
%
%
Another set of modes that the AI PDE is not able to capture 
corresponds to in-phase (IP) oscillating dark soliton stripes 
(hereafter, these will be called IP modes). Nevertheless, one can approximate 
the relevant oscillation frequency as follows. According to the AI analysis 
the oscillatory modes of a {\em single} dark soliton in a 1D trap, are 
characterized by the frequencies~\cite{aipaper2} (see also Ref.~\cite{coles}):
\begin{equation}
\label{eq:IPkn}
\lambda_n=i\,\omega_n = \sqrt{\frac{1}{3}\mu k_n^2-\frac{1}{2}\Omega^2},
\end{equation}
where $k_n= n\pi/L_y$. Then, for in-phase oscillating soliton stripes
and large chemical potentials, 
one may approximate the oscillation frequency of the relevant IP modes 
by Eq.~(\ref{eq:IPkn}).
This is because for such IP modes, the distance between the stripes
remains the same during the evolution and thus the interaction
(depending on their relative distance) effectively does not affect
the motion.
These modes are depicted by the violet curves in Fig.~\ref{DS2spectrum}.
A relevant example concerns the appearance of 
the important stable IP mode, namely the $k_n=0$ mode (see thick green line) 
in the full BdG/GPE spectrum shown in Fig.~\ref{DS2spectrum}. 
The eigenfrequency of this mode corresponds to the oscillation 
frequency Im$(\lambda)=\Omega/\sqrt{2}$ of a single- \cite{buschanglin} 
or multiple-in-phase \cite{ourmarkus3} solitons.  
However, one should not expect that the frequencies of the IP modes in the 
full BdG/GPE will be
identical to $\omega_n$ in Eq.~(\ref{eq:IPkn}): 
this is due to the fact that, 
for the same chemical potential $\mu$, the solitons of 
different $n$ are placed at different positions and hence also have different 
local chemical potentials. This results in the {\em collective} oscillation 
for the multi-soliton stripe state to have a frequency that is slightly larger
than the single-dark soliton prediction~(\ref{eq:IPkn}).
%
%
Here it should be pointed out that the convergence of the full BdG/GPE 
and the $n=0$ IP modes in the TF limit, as shown in Fig.~\ref{DS2spectrum}, 
can simply be understood as follows: 
as $\mu$ increases, the interactions become increasingly short-ranged and, hence, 
the solitons are all pushed towards the center of the trap. Thus, they should all 
oscillate with the same IP frequency ($\Omega/\sqrt{2}$) at the bottom of the well.
We shall see this mode again when we generalize to 3- and 4-stripes in the 
next subsection.
%

%

\begin{figure}[tbp]
\begin{center}
\includegraphics[width=\columnwidth]{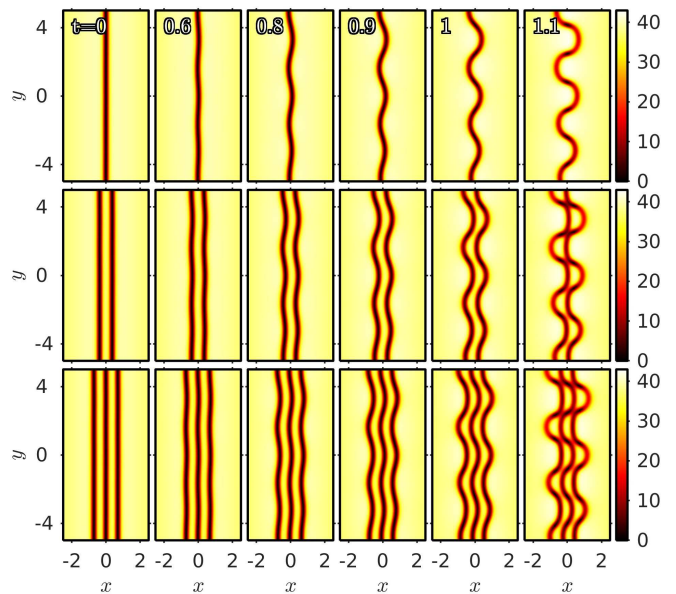}
\caption{(Color online) 
Dynamics for in-phase modes corresponding, from top to bottom,
to the 1-, 2-, and 3-stripe states for $\mu=40$.  
The colorbars on the right-hand side correspond to the density $|u(x,y,t)|^2$.
The system was initialized at $t=0$ with an $N$-stripe state perturbed
by the mode $A\,\cos(k_n y)$ in the transverse direction 
where $k_n=n\pi/L_y$ with $A=10^{-3}$, $n=5$, and $L_y=8$
(only the portion corresponding to $-5\leq y\leq5$ at the
times indicated is shown).
Note that the dynamics, including the growth rate, 
seem to be closely analogous in the different cases. 
}
\label{fig:123DSS_IP}
\end{center}
\end{figure}

Similar to the results for the $n=0$ IP mode, the $n\geq1$ predictions
of Eq.~(\ref{eq:IPkn}) ---see the thick violet curves in Fig.~\ref{DS2spectrum}---
also fare favorably when compared to the full BdG/GPE spectra.
In this case, the lower the order $n$ of the mode, the better the approximation
in Eq.~(\ref{eq:IPkn}) is, and all results asymptotically match as
$\mu\rightarrow\infty$.
It is evident that the modes of the full BdG/GPE analysis
arise in pairs i.e., IP and OOP for the case of 2 stripes.
Similarly, we will see below that the modes arise in triplets
for 3 stripes, groups of 4 for 4 stripes, etc. Among these,
we have confirmed that the lowest growth rate corresponds
to the IP excitation, while the higher growth rate to the OOP one.

An example of the relevant dynamics of in-phase evolution of
multi-soliton stripes is 
depicted in Fig.~\ref{fig:123DSS_IP} for the case of the $n=5$ IP
mode for the 1-, 2-, and 3-stripes and for a 
relatively large chemical potential $\mu=40$. As can be seen from the
figure, the instability evolutionary dynamics
(and thus the associated growth rates) seem to be largely independent
of the number of
stripes. Furthermore, the actual destabilization evolution of the stripes
even in the nonlinear regime (but before the stripes break into pairs
of vortices) seems to also be largely independent of the number of stripes.
Hence, the IP mode excitations behave similarly to the case of a single
stripe.

\begin{figure}[tbp]
\begin{center}
\includegraphics[width=\columnwidth]{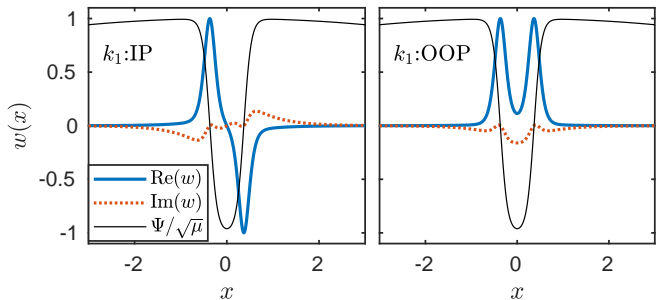}
\caption{(Color online) 
Eigenfunctions corresponding to the in-phase (IP; left) and out-of-phase
(OP; right) unstable modes of the stationary 2-stripe state $\Psi(x)$ 
for $L_y=2$ and $\mu=40$.
Shown are the $k_1$ modes with 
transverse dependence given by $\cos(k_n y)$ with $k_n=n\pi/L_y$.
The eigenmodes corresponding to other values of $n$ are very 
similar (results not shown here).
The thick solid (blue) and dotted (red) lines correspond, respectively,
to the real and imaginary parts of the eigenfunction $w(x)$.
The corresponding normalized steady state, $\Psi(x)/\sqrt{\mu}$,
is depicted by the thin black line.
Note that the real part of the eigenmodes have localized ``humps'' that,
when out-of-phase, produce in-phase motion of the dark solitons and
vice versa. The reason for this apparent contradiction is that the
dark solitons are themselves of opposite phase: in this figure the
left soliton corresponds to a $-\tanh$ while the right one corresponds
to a $+\tanh$.
}
\label{fig:eigenvecs}
\end{center}
\end{figure}

\begin{figure*}[htb]
\begin{center}
\includegraphics[width=0.90\textwidth]{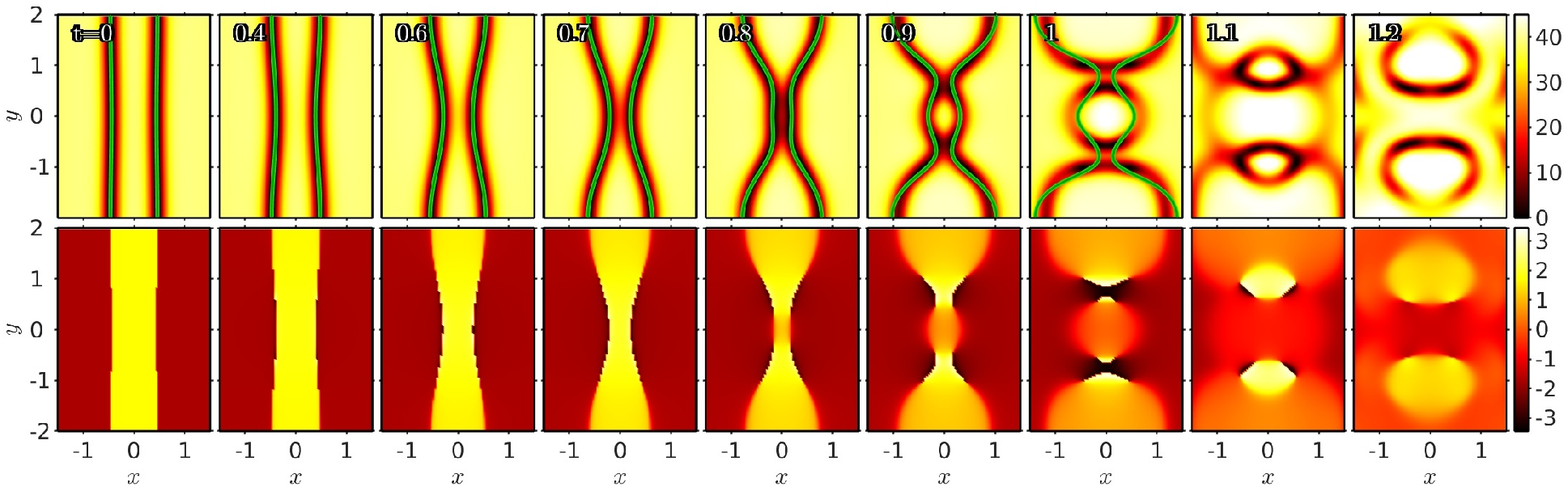}
\\[2.0ex]
\,
\includegraphics[width=0.905\textwidth]{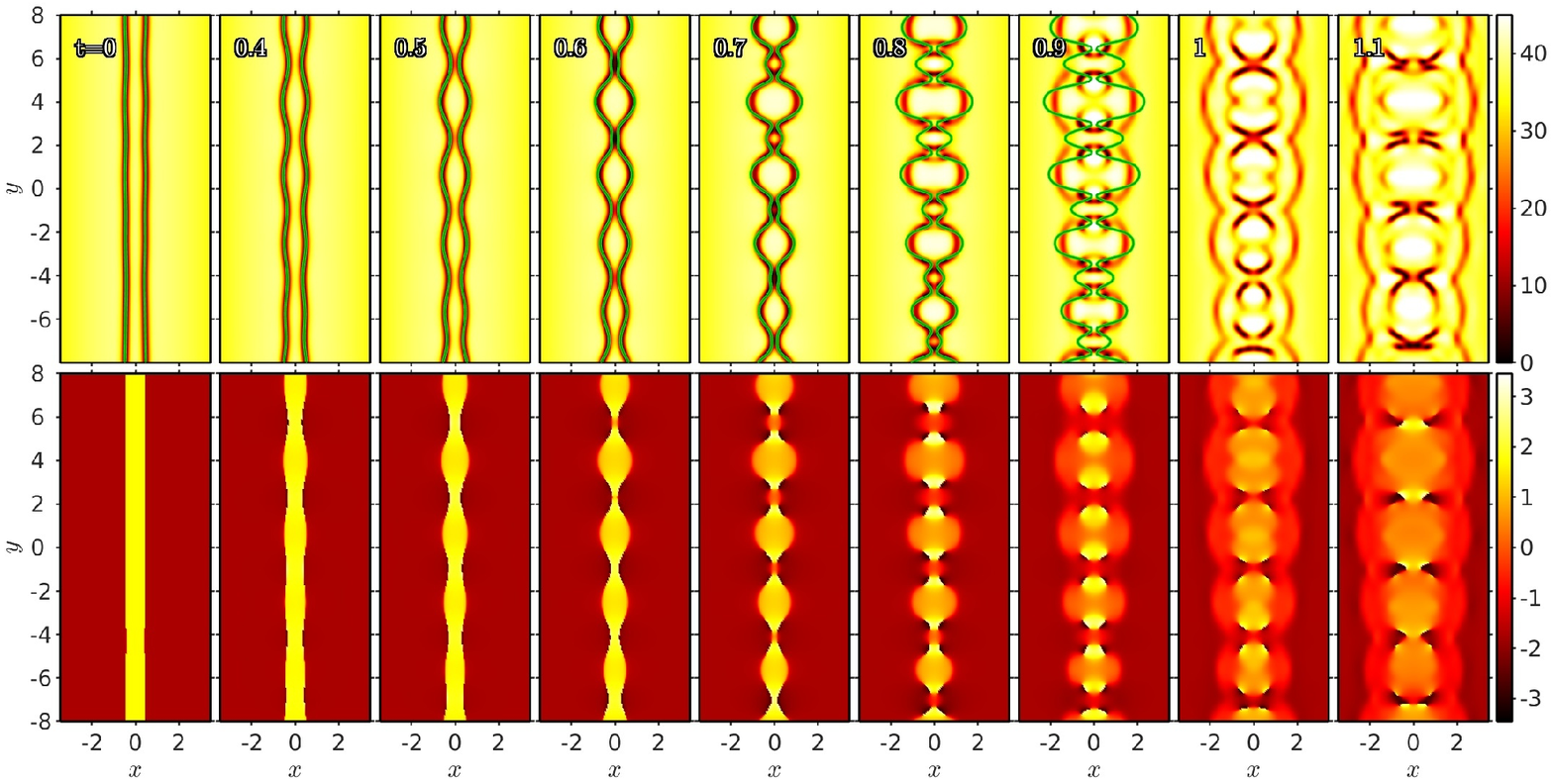}
\caption{(Color online) 
Dynamical destabilization and collision of two dark soliton stripes.
The background colormap corresponds to full GPE numerics while
the green overlaid curves correspond to the AI reduction
The corresponding systems are initialized symmetrically with 
the right dark soliton stripe at 
$x_0(y) = \xi_0 + \sum_{n=1}^{\nu}\varepsilon_n\,\sin(2\pi ny/L_y+\varphi_n)$
with $\varepsilon_n=0.01$, $\varphi_n=(n-1)L_y\pi/10$, 
$\xi_0=0.45$ (i.e., close to the steady state equilibrium 
$x_0^{(\rm{eq})}=0.368$), and $\mu=40$.
The top two rows of panels correspond to a single-mode
perturbation ($\nu=1$) with $L_y=2$ while the bottom two rows
of panels correspond to a perturbation containing five modes ($\nu=5$)
and $L_y=8$.
Within each set of panels the top and bottom rows (and their colorbars) correspond
to the density ($|u(x,y,t)|^2$) and phase ($-\pi$ to $\pi$) of the field at the
indicated times.  
%
}
\label{fig:dyn1}
\end{center}
\end{figure*}

Finally, as concerns the spectra, it is worth mentioning that 
there exist other eigenmodes corresponding to the background
(rather than the dark soliton stripes).
For instance, there exist imaginary eigenvalues that appear in pairs 
(starting at the linear limit) and that monotonically increase as a
function of $\mu$. These eigenvalue pairs seem to quickly approach
(asymptotically) each other as $\mu$ is increased (for the $L_y=4$
case they visually coalesce around $\mu\approx 20$; see
the relevant monotonically increasing blue curves in the bottom
panel of Fig.~\ref{DS2spectrum}).
Each of these eigenmode pairs corresponds to undulations of the background's 
edges where the left and right ends oscillate with progressively
higher wavenumbers, the higher the mode frequency.
These oscillations are in- or out-of-phase between the two edges
(at opposite $x$'s), yet as
the chemical potential increases, and so does the separation between
the edges, the former and latter tend to oscillate with the same frequency.
We do not describe further these modes as they only pertain to benign
(i.e., purely oscillatory) perturbations
away from the dark soliton stripes.

With the above observations in mind, let us now summarize a 
number of key features of the multi-stripe spectra: 
\begin{itemize}
\item[(1)] Our theory correctly captures with good accuracy the {\it stable} $n=0$ 
out-of-phase (OOP) oscillation mode (see analysis in the end of Sec.~\ref{theory}) of the two stripes, 
in line with the pertinent 1D theory~\cite{coles}. 

\item[(2)] The oscillation frequency of the stable $n=0$ in-phase (IP) mode for the 2-stripe state 
can be well approximated by the frequency for the 1-stripe state in the trap; 
this is confirmed in the full BdG/GPE spectrum.
\item[(3)] Our theory detailed above in Sec.~\ref{theory}
only captures the OOP modes. The IP modes appear to be 
independent of the interaction, and hence essentially
equivalent to the single-stripe results of 
Ref.~\cite{aipaper2} in the TF limit. This also refers to the 3- and 4-stripe case 
(see below).

\item[(4)] Each of the finite $k_n$ transverse wavenumber modes,
compared with the single-stripe case, splits into two. This has a 
natural explanation as the number of stripe degrees of freedom has been
doubled. In a similar vein, there are three sets of branches for the 3-stripe case, 
and four branches for the 4-stripe case (see below).  
In Fig.~\ref{fig:eigenvecs} we depict the longitudinal dependence
of the IP and OOP modes extracted from the full BdG/GPE. 
As expected, for two stripes there are two ``normal'' modes for each
$k_n$ corresponding to the IP and OOP ones.
\item[(5)] Despite the increase in the multiplicity of the normal modes, 
different normal modes of the same $k_n$ mode have similar growth rates, 
also similar to those of a single dark soliton stripe. Therefore, the 
increase of the number of solitons creates more instabilities, but does
not substantially increase the instability growth rate.
Nevertheless, as we will show below, for large $\mu$ there is a monotonic
(albeit weak) increase of the growth rate of the instabilities with
the stripe number.
It also appears 
that the branches of the same unstable $k_n$ mode for the two stripes 
converge together in the TF limit. 
\item[(6)] There is also one interesting difference near the linear limit. 
In contrast to the single stripe, which has a narrow fully stable regime 
near the linear limit, many stripes (including the following
cases of 3- and 4-stripes)
studied here all become unstable right from the linear limit. 
Note the small ``bumps'' in $\lambda_r$ near the linear limits
(cf.~inset in the top panel in Fig.~\ref{DS2spectrum}). 
This instability is analogous to the one observed in the corresponding
quasi-1D~\cite{ourmarkus3} and 1D~\cite{coles} cases, arising from 
opposite energy mode collision and
being associated with complex eigenvalues.
However, before the latter is stabilized
(as in 1D), transverse modes start yielding unstable growth
(via real eigenvalues). This causes
the multi-stripe configurations to be susceptible to instability for
{\it all} the values of chemical potentials considered herein.
%
%
\end{itemize}


\begin{figure}[tbp]
\begin{center}
\includegraphics[width=0.90\columnwidth]{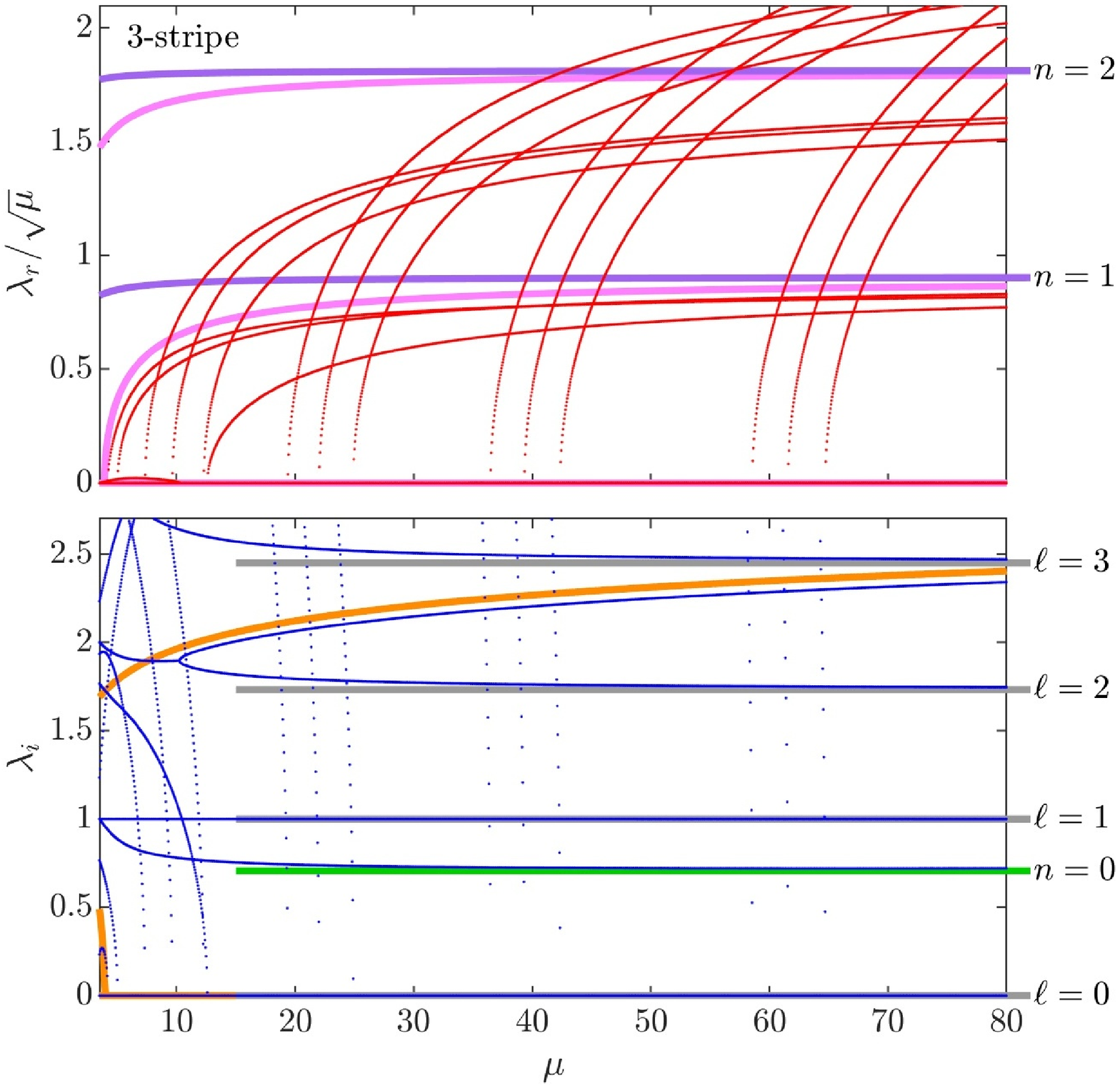}
\includegraphics[width=0.90\columnwidth]{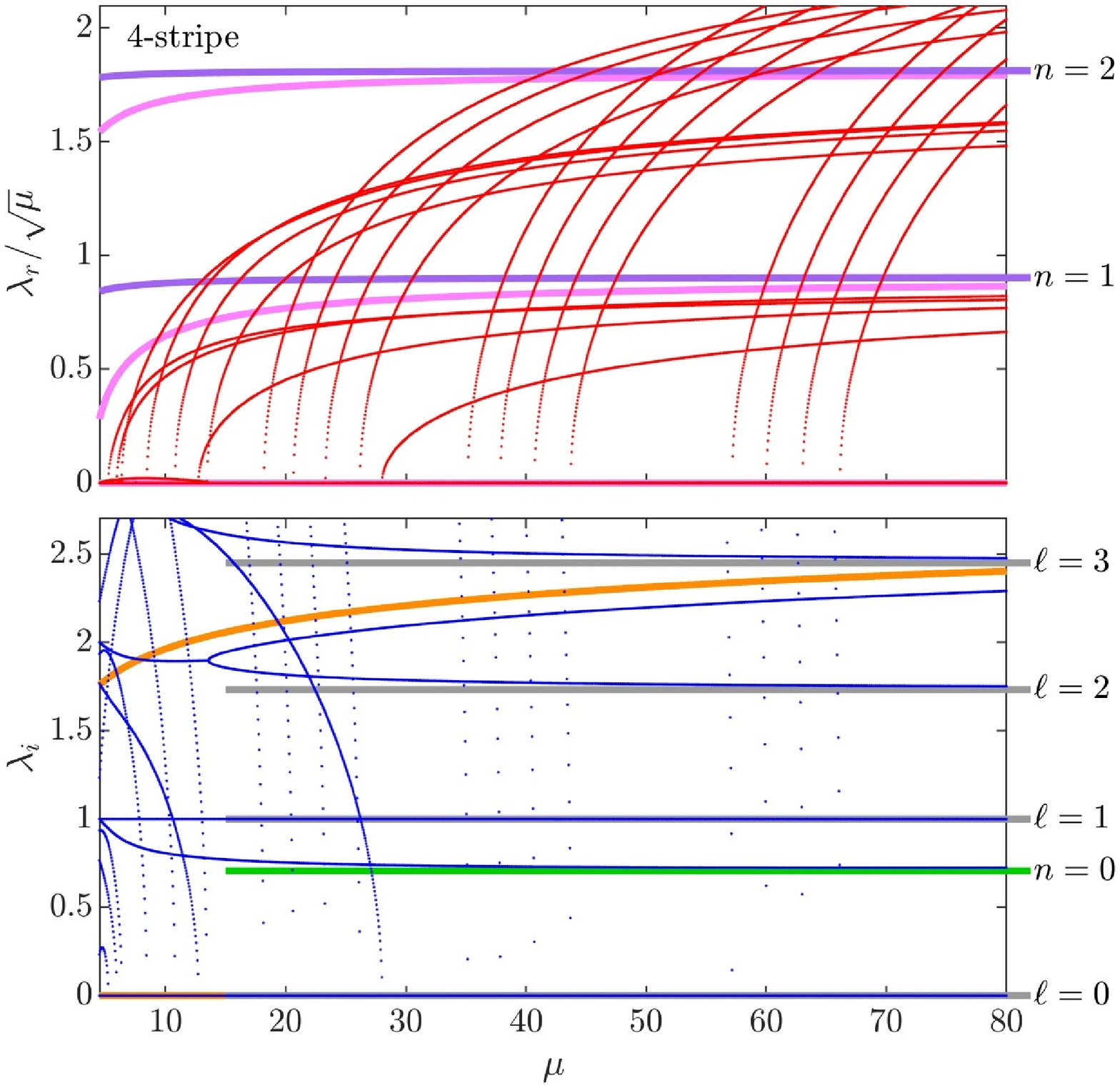}
\caption{(Color online) 
Similar to Fig.~\ref{DS2spectrum} (for 2-stripes), but now for 
the 3- and 4-stripe cases, for $L_x=16$ and $L_y=2$. 
}
\label{DS34spectrum}
\end{center}
\end{figure}

The results presented above indicate that the AI PDE is able to approximate
very well the (linear OOP) modes that it is designed to capture. However, our
ultimate goal with the AI method is not only to obtain the BdG modes,
but more importantly to obtain a reduced AI PDE that is also able to capture
the (linear and) nonlinear dynamics of the stripe dynamics.
We therefore now compare the dynamics of the full GPE and of the AI PDE. 
We initialize typical symmetric stripes of the following form 
\begin{equation}
x_0(y,t=0) = \pm( \xi_0 + p(y)),
\end{equation}
where $\xi_0$ is the initial location of the stripes and $p(y)$ 
is a perturbation to accelerate the destabilization dynamics.
We depict in Fig.~\ref{fig:dyn1} two typical cases that compare
the full GPE dynamics (see background colormap) and the AI PDE dynamics
(see overlaid green curves).
The first case (see top set of panels) corresponds to an initial position
close to stationary equilibrium $x_0^{(\rm{eq})}$ perturbed by a single
mode ($n=1$ for $L_y=2$) with amplitude 0.01. 
The second case (see bottom set of panels) corresponds to the same initial 
position of the stripe but now perturbed by five different modes (see caption 
for more details); see figure caption for details on the perturbation 
$p(y)$ used in these two cases.
As can be noticed from the figure, for both cases, the AI PDE is able
to capture (a) the initial growth of the perturbation (in accordance with
our previous results on the BdG spectra), (b) the strong interactions between
the stripes that include collision and bounce-back, and importantly
(c) even some of the
nonlinear stripe dynamics before the stripes finally break into vortices.
It is remarkable that indeed the AI PDE is able to capture the full GPE
dynamics even when the stripes are interacting quite strongly.


\subsection{3- and 4-stripe states}

\begin{figure}[tbp]
\begin{center}
\includegraphics[width=\columnwidth]{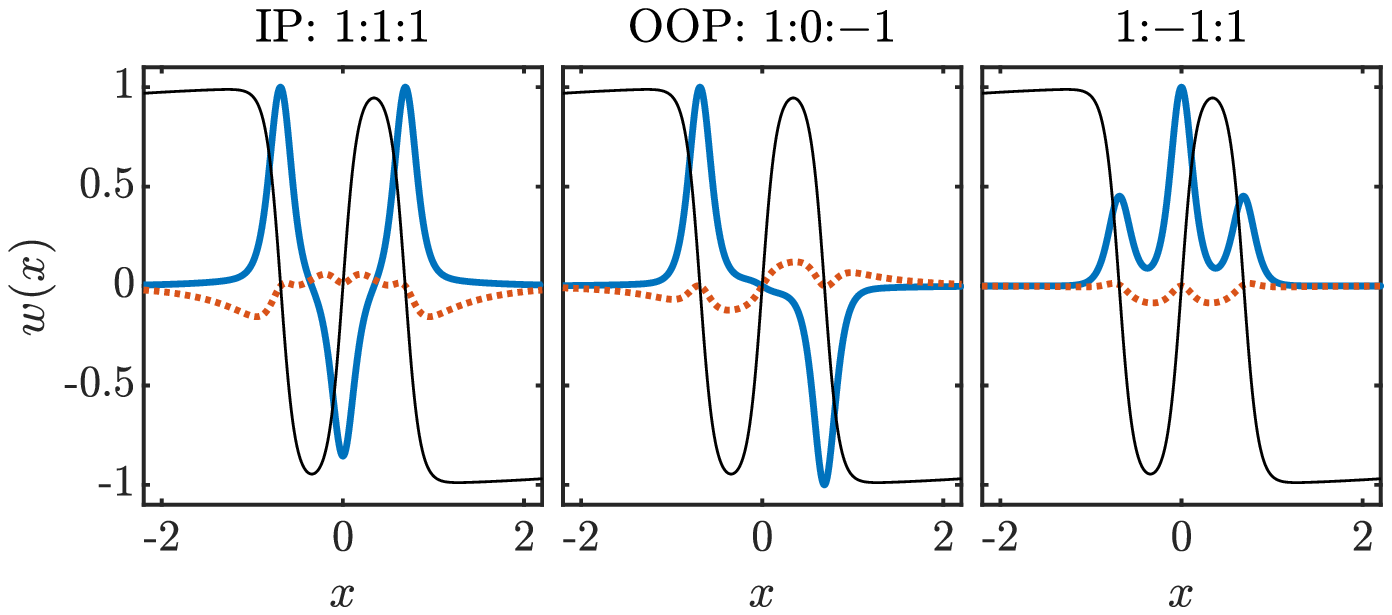}
\\[2.0ex]
\includegraphics[width=\columnwidth]{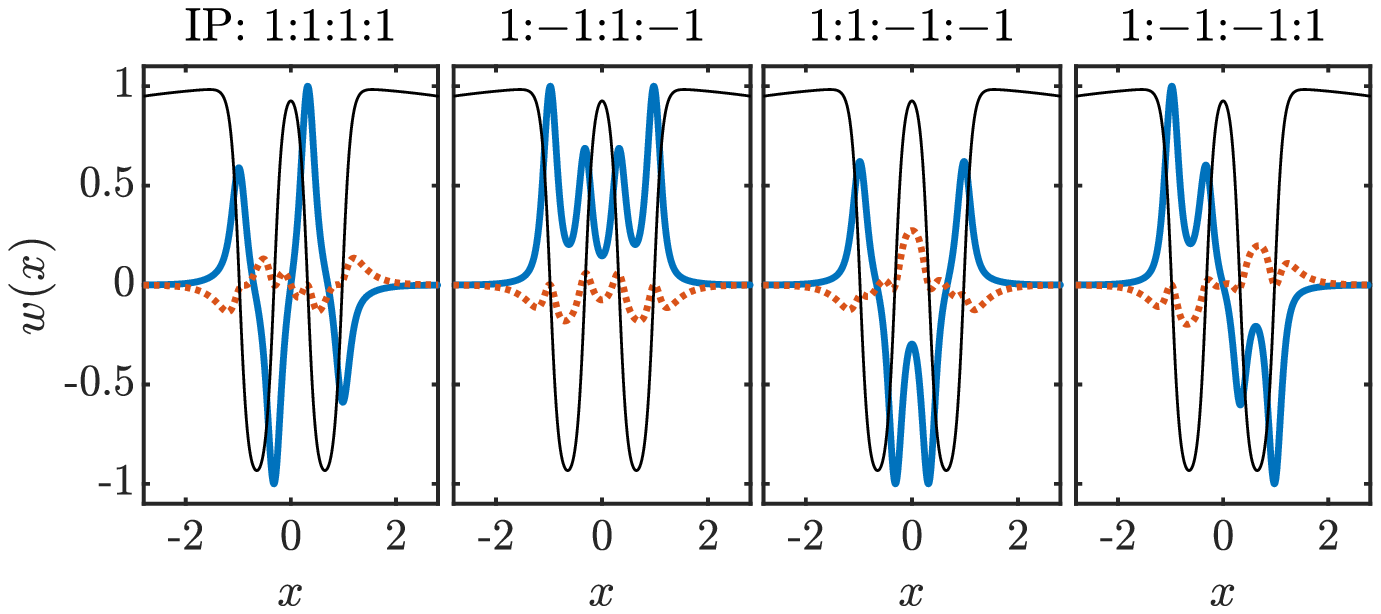}
\caption{(Color online) 
Same as Fig.~\ref{fig:eigenvecs} but for the 3-stripe (top) 
and 4-stripe (bottom) states. 
Parameters correspond to $L_y=2$ and $\mu=40$ for the
transverse modes involving the motion of the dark soliton stripes.
The notation $\pm1:\pm1:\pm1$ is used to denote a normal mode
that has a displacement from the steady state of the first,
second, and third dark soliton in the positive (+) or negative
($-$) direction. The same notation is used for the 4-stripe case.
}
\label{fig:eigenvecs34}
\end{center}
\end{figure}

\begin{figure}[tbp]
\begin{center}
\includegraphics[width=0.75\columnwidth]{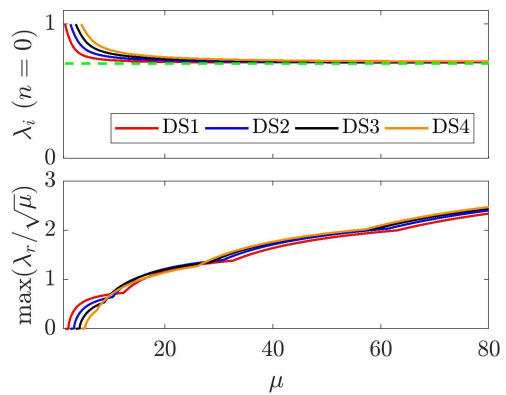}
\caption{(Color online) 
Top panel: Comparison of the fully in-phase oscillation mode $(n=0)$ frequency 
for different numbers of dark soliton stripes using the same $L_y=2$. 
DS$N$ denotes the $N$-stripe state.
Note that they all converge to the 1D single dark soliton results 
$\Omega/\sqrt{2}$ corresponding to the IP oscillation of the $N$-stripes.
%
Bottom panel: The most unstable mode growth rate for different numbers of dark soliton 
stripes. Note there is an interesting crossover behavior. The growth rate 
is larger for the 1-stripe state near the linear limit but in the TF limit, 
more dark solitons are more unstable. Moreover, in the latter limit
there is a (weak, yet) monotonic increase of the growth rate with $N$.
}
\label{Statistics}
\end{center}
\end{figure}

In this section, we consider the cases of 3- and 4-stripe states, 
and study their spectra and dynamics. The dependence of the real and imaginary parts 
of the eigenvalues on the chemical potential is shown in Fig.~\ref{DS34spectrum}
for both cases. 
Besides the presence of the same background modes as the ones describe
above for the 2-stripe case, we can clearly see the multiplicity of 
the inter-stripe vibrational
modes.  The 3-stripe case has three modes while the 
4-stripe case has four.
As explained above, this multiplicity stems from the different normal
modes of vibration of the $N$-stripe solution. Indeed, we have
extracted the longitudinal dependence of the normal modes for $n=2$
as depicted in Fig.~\ref{fig:eigenvecs34}. As can be observed from the
figure, the 3- and 4-stripe cases have, respectively,
three and four normal modes.

It is also evident, more so in the 3- and 4-stripe case
when compared to the 2-stripe case, that in comparison, e.g.,
with Fig.~2 of Ref.~\cite{aipaper2},
the higher $N$ is, the more unstable the
corresponding state becomes in the TF limit. 
This makes intuitive
sense, as the differential repulsion of the stripes at different
locations (in the presence of perturbations) can be expected to lead to
enhancement of the undulations and ultimately of the instability
growth rate. 
This is more concretely quantified in Fig.~\ref{Statistics}. The top panel
clearly shows
that the IP modes converge, in the TF limit,
to the same frequency of IP oscillations for the different stripes.
%
On the other hand, the bottom panel shows, interestingly, that 
the single-stripe state is the one that bifurcates into instability
the earliest.
%
Nevertheless, and going towards the analytically tractable
limit of large $\mu$, we observe that there exists
a crossover. As a result, more
soliton stripes lead to higher ---but only slightly higher--- 
growth rates. Nevertheless, the scales of the growth rates of
different numbers of stripes remain
quite proximal.

\begin{figure}[htb]
\begin{center}
\includegraphics[width=0.90\columnwidth]{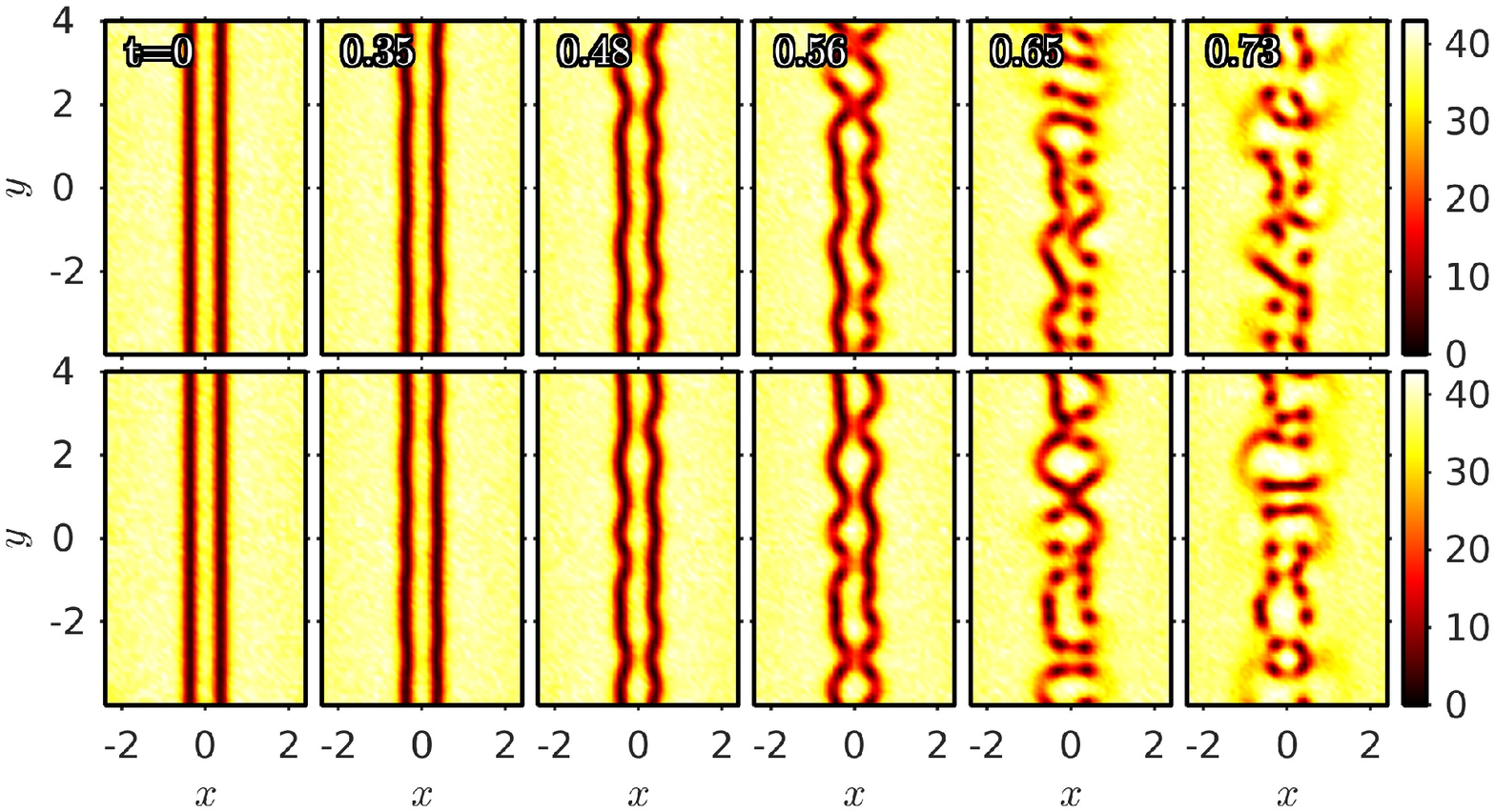}
\\[0.5ex]
\includegraphics[width=0.90\columnwidth]{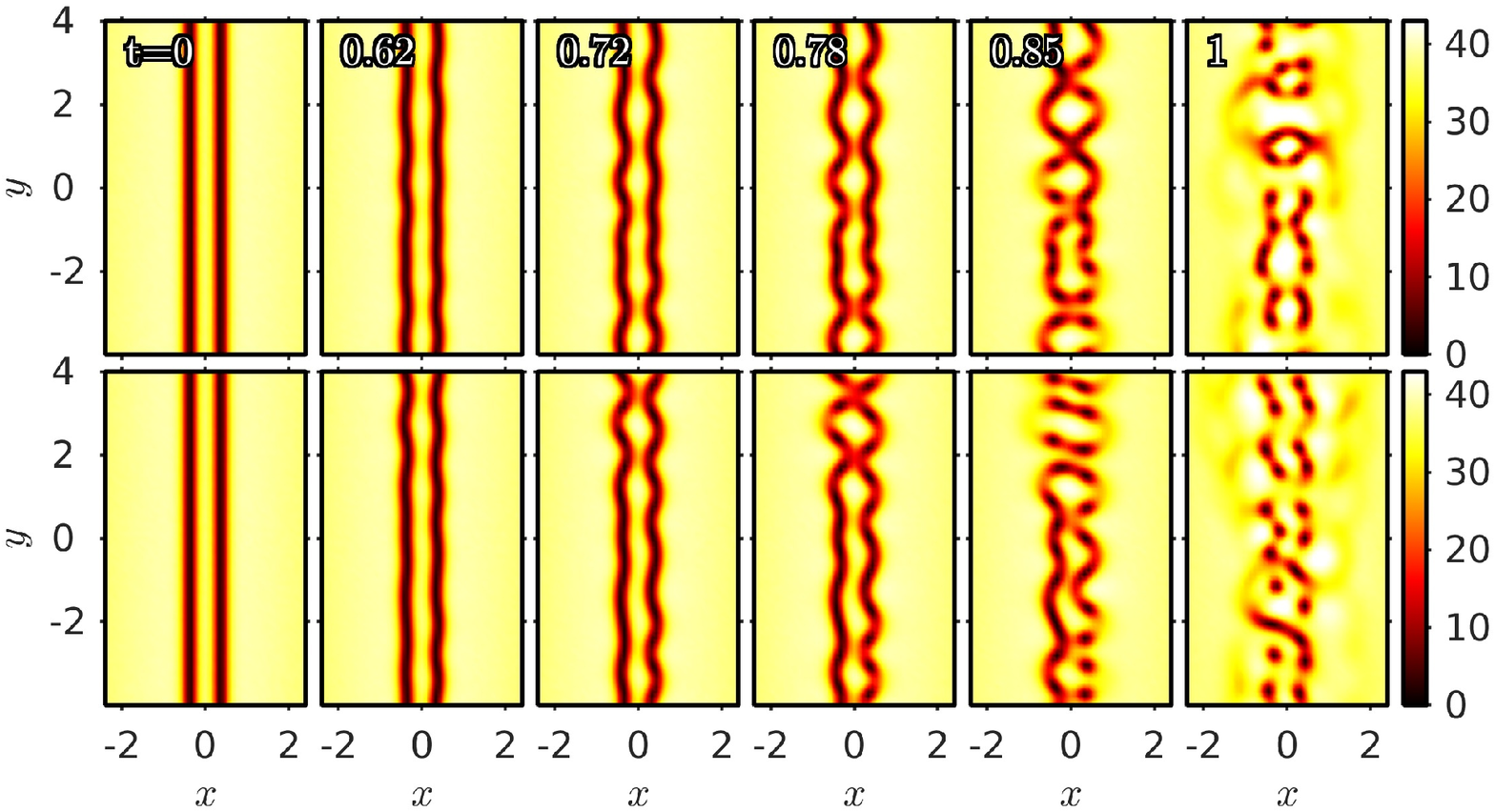}
\\[0.5ex]
\includegraphics[width=0.90\columnwidth]{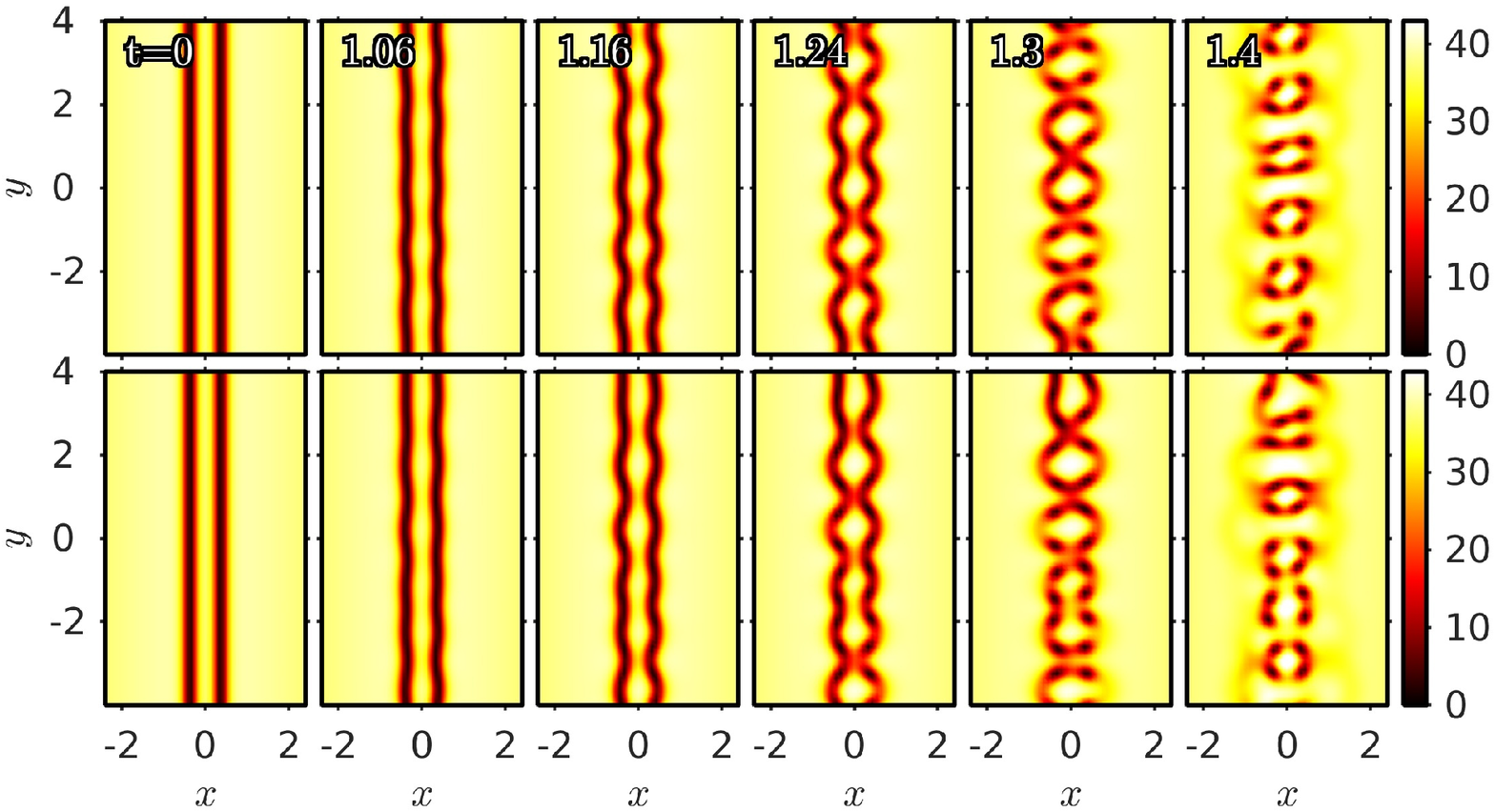}
\\[0.5ex]
\includegraphics[width=0.90\columnwidth]{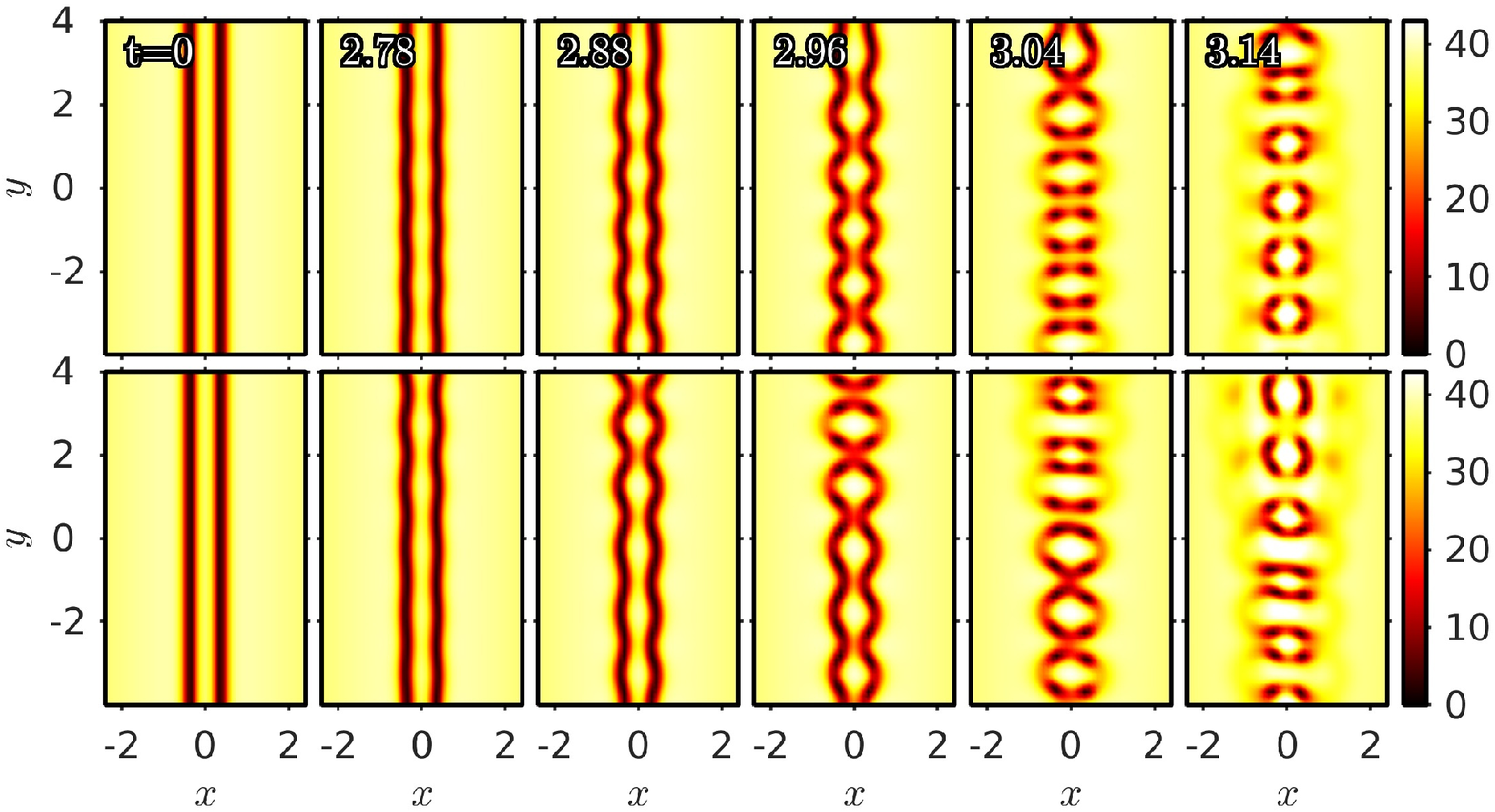}
\caption{(Color online) 
Dynamical destabilization of 2-stripe states
using different small random perturbations.
Each set of two rows of panels corresponds, from top to bottom, to 
two different realizations for a random perturbation of amplitude
$10^{-1}$,
$10^{-2}$,
$10^{-4}$, 
and 
$10^{-12}$, 
respectively.
}
\label{fig:D2rand}
\end{center}
\end{figure}

\begin{figure*}[htb]
\begin{center}
\includegraphics[width=0.9\textwidth]{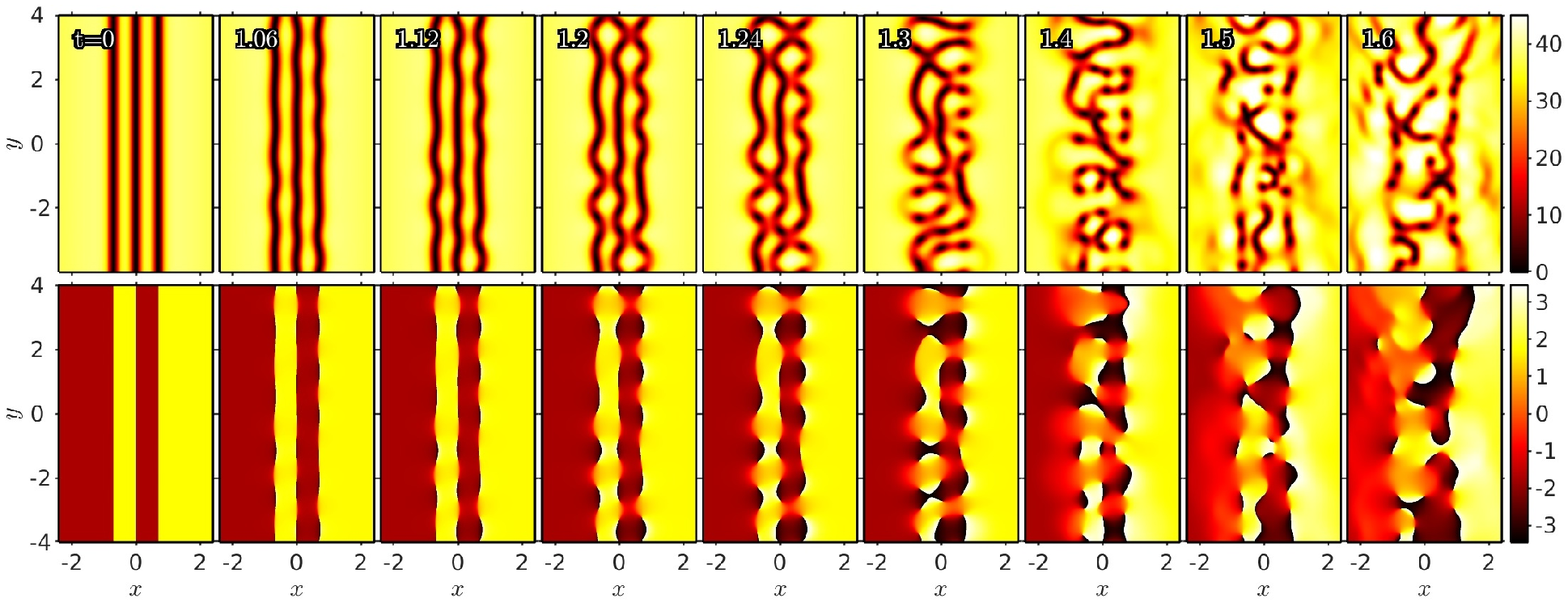}
\includegraphics[width=0.9\textwidth]{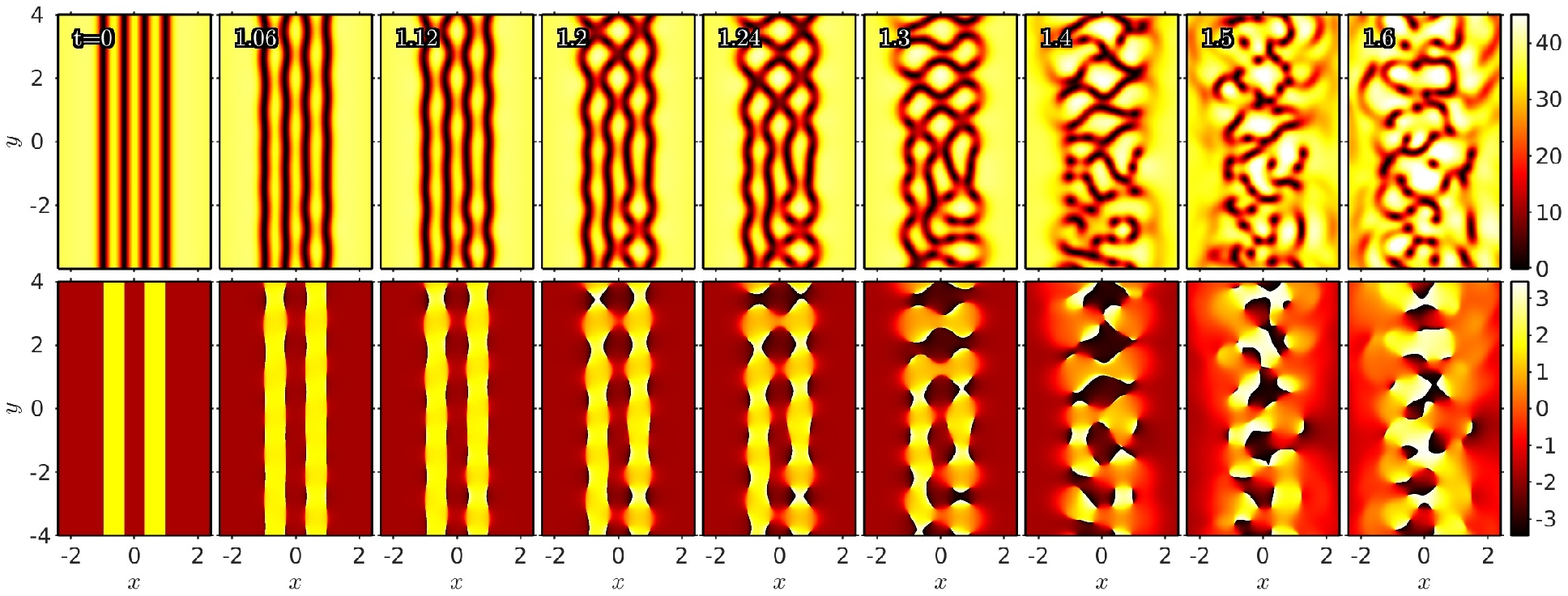}
\caption{(Color online) 
Dynamical destabilization of the 3- and 4-stripe states
as obtained from the full GPE. In these cases, we have added a small 
($10^{-4}$ amplitude) random noise. 
Both states, following the spectral picture, break with the most unstable
mode, and  give rise to numerous vortex pairs.
Note that despite the existence 
of more unstable modes, the maximum decay rates are about the same 
for different number of stripes. 
}
\label{D3}
\end{center}
\end{figure*}

Finally, we have performed simulations to study the dynamics
when random noise is added to the stationary state.
In this case we do not compare with the AI PDE dynamics as
our analytical characterization captures the OOP modes (in terms
of their nonlinear dynamics) but not the IP ones. Given that
random perturbations excite both, a more elaborate and less straightforwardly
tractable AI approach without the assumption of symmetric center positions
would be needed for comparison here.
In Fig.~\ref{fig:D2rand} we depict the dynamics ensuing from the
2-stripe state perturbed by a small, uniformly distributed random perturbation.
For each perturbation size we show two typical runs and, from top to
bottom, we depict the cases for random perturbations of 
diminishing amplitude. Surprisingly, as the perturbation is set to
smaller values, the full GPE dynamics tends to a {\em symmetric OOP}
configuration for the two stripes. 

In Fig.~\ref{D3} we depict the dynamical destabilization for the
3- and 4-stripe cases.
The BdG spectra depicted in Fig.~\ref{DS34spectrum}, computed for
$L_y=2$ suggest that the dominant unstable mode for $\mu=40$ 
should be the $n=3$ one which corresponds to the $n=6$ mode
in Fig.~\ref{D3} as in this latter case we used $L_y=4$.
The dynamics indeed follows this prediction until
each stripe breaks into vortex pairs.
It is, arguably, not entirely straightforward to define a precise breaking
time, as this is a continuous process, but both states start to bend
around $t=1$ and therefore the two states break at about
the same time scale, as expected from the spectra. 
It should be noted that for the 3-stripe case, the AI approach is
conceptually straightforward to construct, generalizing the 
energy functional of Eq.~(\ref{dark5}), as for OOP configurations
the dark soliton in the middle is centered at $x=0$ and the
relevant center positions are $-x_0$, $0$ and $x_0$.
However, the resulting expressions are particularly tedious and 
hence we do not attempt to give them here.

We now comment on the possibility 
of generating quantum turbulence from the transverse-instability-induced  
dynamics of multiple dark soliton stripe states. This scenario was considered 
and investigated in Ref.~\cite{tsubota} for an initial configuration different from 
that used in this work, namely for a square grid of many dark soliton stripes in a 
spatially uniform condensate. In our case, Fig.~\ref{D3} (see panels corresponding 
to $t=1.6$) suggests that the instability-induced generation of vorticity could also 
lead to quantum turbulence. Nevertheless, the state of quantum turbulence that 
is induced by the decay of a series of dark soliton stripes is likely to lack the 
vortex clustering, as well as the 
statistical signatures of classical and quantum two-dimensional
turbulence~\cite{Bra2012.PRX2.041001,Ree2014.PRA89.053631,Bil2014.PRL112.145301}.
Nonetheless, weak correlations and small clusters can build after some 
vortex-antivortex annihilation~\cite{Sim2014.PRL113.165302}, which we surmise is one of
the more interesting effects that could be observed for this type of turbulent 
state. In any case, a pertinent systematic study of such effects is beyond the 
scope of this work.

\section{Conclusions \& Future Work}
\label{cc}

In the present work, we have extended considerations of the solitonic
stripes as filaments to the realm of multi-stripes, taking into
consideration their pairwise interactions. We have seen how this
allows one to evaluate the equilibrium position of multi-stripe
states. Perhaps more importantly, this also enables the consideration
of the linearized eigenmodes around such an equilibrium. These
modes can be partitioned into in-phase and out-of-phase ones. The in-phase
ones are similar to 
the single-stripe modes. On the other hand,
the out-of-phase ones introduce additional growth modes of the transverse 
instability. Despite the larger number of instabilities, the maximum growth 
rates remain comparable to that of a single stripe, although we have found a
weak monotonic dependence thereof on the the number of stripes $N$
in the large chemical potential limit.
In addition to this linearization picture, we have explored the full
dynamics of the two stripes, always in good agreement with the
filament (AI PDE)
method results which consider each of the stripes as a reduced PDE
for the stripe center $x_0(y,t)$. We have extended the numerical
consideration of such stripe interaction scenarios to the case
of 3- and  4-stripe settings, obtaining a natural generalization
of the two-stripe results.

This effort paves the way for a number of future possibilities.
One of the most intriguing ones, in line with the experimental
thesis results of Ref.~\cite{kali}, is to examine the interaction of
a quasi-1D pattern (like the stripe) and a genuinely 2D pattern,
like the vortex. This has been associated with a nonlinear variant
of the famous Aharonov-Bohm effect in Refs.~\cite{neshev1,neshev2}.
Yet, it has not been systematically explored at the level of a
filament theory as the one presented herein, which could shed
quantitative light in the relevant dynamics. Moreover, this
is an especially appealing problem at the interface of dimensionalities
and at the interface between differential and integral equations
(preliminary calculations suggest that the vortex has a distributed
effect on the stripe, while the stripe has an integrated effect
on the motion of the vortex).  A version of this problem that
could be radially symmetric and hence simpler to tackle could
be that of a ring dark soliton with a vortex sitting in its center.
One can also go beyond 2D settings and consider effective
PDEs for 1D filaments such as vortex rings embedded in 3D
space as in the recent works of Refs.~\cite{ruban2,ticknew} and then
attempt to generalize these incorporating ring-ring interaction
to account for multi-vortex-ring settings~\cite{leapfrog2,leapfrog3,ourrecent}.
These directions will be considered
in future studies.

\acknowledgments 

W.W.~acknowledges support from the Swedish Research Council Grant 
No.~642-2013-7837 and Goran Gustafsson Foundation for Research in
Natural Sciences and Medicine.
P.G.K.~gratefully acknowledges the support of
NSF-PHY-1602994, as well as from  the Greek Diaspora
Fellowship Program. 
R.C.G.~gratefully acknowledges the support of NSF-PHY-1603058.
B.P.A.~gratefully acknowledges the support of NSF-PHY-1607243.

\end{document}